\documentclass[aps,twocolumn,nopacs,amsmath,amssymb]{revtex4}

\usepackage{graphicx}
\usepackage[british,UKenglish,USenglish,english,american]{babel}
\usepackage{bm}
\usepackage{amsmath}
\usepackage{amssymb}
\usepackage{float}
\usepackage{xcolor,graphicx}
\usepackage{mathrsfs} 

\begin{document}
\title{Magneto-optical effects in the Landau level manifold of 2D lattices  with spin-orbit interaction}
\author{Muzamil Shah}
\author{Muhammad Sabieh Anwar}
\email{sabieh@lums.edu.pk}
\affiliation{Laboratory for Quantum Technologies, Department of Physics, Syed Babar Ali School of Science and Engineering, Lahore University of Management Sciences (LUMS), Opposite Sector U, D.H.A., Lahore 54792, Pakistan}
\date{\today}

\begin{abstract}
 Silicene is a competitive and promising 2D material, possessing interesting topological, electronic and optical properties.  The presence of strong spin orbit interaction in silicene and its analogues, germanene and tinene, leads to the opening of a gap in the energy spectrum and spin-splitting of the bands in each valley. Building upon prior work we discuss a general method to determine the magneto-optic response of silicene when a Gaussian beam is incident on silicene grown on a dielectric substrate in the presence of a static magnetic field. We use a semiclassical treatment to describe the Faraday rotation (FR) and Magneto-optical Kerr effect (MOKE). The response can be modulated both electrically and magnetically. We derive analytic expressions for valley and spin polarized FR and MOKE for arbitrary polarization of incident light in the terahertz regime. We demonstrate that large FR and MOKE can be achieved by tuning the electric field, magnetic fields and chemical potential in these fascinating 2D materials.
\end{abstract}
\maketitle

\section{Introduction}
Monolayer graphene has garnered immense interest from a large global community of researchers. This is primarily due to its unique electronic and optical properties \cite{Novoselov2004} derived from its exotic electronic structure. For example graphene possesses gapless Dirac-type band structure \cite{Tombros2007}, high carrier mobilities and universal broadband optical conductivities (due to inter band transitions) \cite{Koppens2011}. Due to its fascinating optical properties, graphene is also considered to be a promising material for photonic and optoelectronic applications in the terahertz (THz) to mid-infrared ranges. For example, Faraday and Kerr rotations are non-reciprocal magneto-optic (MO) effects, in which the polarization of a plane wave is rotated when linearly polarized light is respectively transmitted or reflected from a transparent medium in the presence of static uniform perpendicular magnetic field $B$. Both of these effects originate from the breaking of time reversal symmetry by an external applied magnetic field. Graphene exhibits an exceptionally large Faraday and Kerr rotation in the THz region and therefore is considered a futuristic candidate for non-reciprocal tunable devices \cite{Crassee2011,Ferreira2011,Sounas2011}. The magnitude of FR is about $6^{\circ}$ in a field of strength 7 T. Unfortunately, the FR and magneto-optic Kerr effects (MOKE) observed in a single layer graphene sheet exist only at low frequencies ($<$ 3 THz) and that too in the presence of large magnetic fields.

Graphene shares analogous properties with a large range of 2D quantum materials \cite{Ahmed2017}. For example, recently, transition metal dichalcogenides (TMDC) have attracted a lot of attention due to their novelty \cite{Ahmed2017,Mak2014}. TMDC's have the formula MX$_2$, where M is a transition-metal atom (Mo, W, V, etc) and $X$ is a chalcogen atom (S, Se, or Te). TMDC's are of particular interest because they possess a valley degree of freedom and exhibit large band gaps due to SOI \cite{Catarina2019}. These interesting spin-valley structures make TMDC's highly attractive condidates for spintronic, valleytronic \cite{Mak2014,Catarina2019,Suzuki2014,Cazalilla2014} and optoelectronic devices \cite{Pospischil2014,Wu2015}.

The discovery of 2D materials has also stimulated growing interest in silicene \cite{Guzman2007}, the silicon analogue of graphene. Stable silicene can be experimentally synthesized \cite{Cahangirov2009}. There are many electronic and physical similarities between graphene and silicene as both are found in the same group of the periodic table. The major difference is that silicene has a large SOI with an electrically tunable band gap. Just like silicene, germanene and tinene also possess stable honeycomb lattice structures \cite{Cahangirov2009,Cai2015}. Due to the relatively large SOI, these materials haved buckled structures, providing a mass to the otherwise massless Dirac fermions. In silicene \cite{Ezawa2012}, germanene \cite{Liu2011} and tinene \cite{Xu2013}, the values of $\Delta_{so}$ have been predicted to lie in the range $1.55\textendash7.9$ meV, $24\textendash93$ meV, and 100 meV  respectively. Subsequently, the interaction of an external electric field with silicene, germanene and tinene-substrate system renders the Dirac mass controllable at the $K$ and $K'$ points, which leads to various topological phase transitions \cite{Ezawa2013}.

In addition to charge and spin, which are intrinsic degrees of freedom, Dirac electrons have another degrees of freedom called the valley \cite{Schaibley2016,Xiao2007,Yao2008}. The valley can also be used to encode and process information, this is the now burgeoning field of valleytronics \cite{Schaibley2016}. A promising platform for valleytronics is provided by silicene.  Due to spin and valley polarized responses, silicene also offers the possibility to realize novel tuneable MO devices \cite{Xiao2012,Zeng2012}.

The possibility of dynamic adaptability of silicene's electronic structure via electric and magnetic fields makes it favorable for tuneable THz applications. However, the two most important MO responses namely FR and MOKE of monolayer silicene and the wider class of Dirac materials deserves a rigorous exploration. The purpose of this work is to study FR and MOKE in these 2D lattices. Subsequently, the magnetic field dependent MO effects can be directly utilized for magnetic field sensing and optical modulation \cite{Mihailovic2013,Yoshino2003,Dawson1995}. In addition to FR and MOKE, in this work, we also investigate the dependence of these MO effects on the incident angle, polarization state, chemical potential and temperature.

\subsection{System Hamiltonian and MO conductivities}
The starting point for the derivation of Faraday and Kerr rotations, and ellipticities, is the understanding of the energy levels manifold. This aspect has been thoroughly presented by several authors \cite{Tahir2013,Tabert2013,Ezawa12012} and in this section, we only \emph{reproduce} prior results. The low-energy physics of silicene, germanene and tinene is adequately approximated by a simple nearest-neighbor tight-binding Hamiltonian
\begin{equation}\label{a1}
\hat{H}_{\xi \sigma}=\hbar v_{F}(\xi k_x \hat{\tau}_{x}+k_y \hat{\tau}_{y})-\frac{1}{2}\xi  \Delta_{so}\hat{\sigma}_{z}\hat{\tau}_{z}+\frac{1}{2}\Delta_{z}\hat{\tau}_{z}\cdot
\end{equation}
This Hamiltonian is generalized by Cysne et al. \cite{Cysne2018} to include Rashba and valley Zeeman SOI. 
The first term in Eq.~(\ref{a1}) is the usual low-energy graphene-like Hamiltonian for describing massless Dirac fermions, $k_{x,y}$ are their crystal momentums and $v_{F}$ is their Fermi velocity. The parameter $\xi=\pm 1$ corresponds to the valleys ($K$ and $K^{'}$) in momentum space and the vector operators $\vec{\tau}=(\hat{\tau}_{x},\hat{\tau}_{y},\hat{\tau}_{z})$ and $\vec{\sigma}=(\hat{\sigma}_{x},\hat{\sigma}_{y},\hat{\sigma}_{z})$ respectively represent Pauli matrices of the lattice pseudo spin and real spin degrees of freedom.  The second term in the Hamiltonian captures intrinsic spin-orbit coupling with a band gap of $\Delta_{so}$, whereas in the final term, $\Delta_z=aE_z$ is responsible for breaking the $A$, $B$ sublattice inversion symmetry, $E_{z}$ being an electric field normal to the plane of atoms and $a$ being the lattice constant. For Landau level (LL) quantization, we apply a static uniform magnetic field $B$ perpendicular to this plane.
Introducing the Landau gauge for the magnetic vector potential $A = (-yB, 0,0)$, and diagonalizing the  Hamiltonian, we obtain the eigenvalues \cite{Tabert2013},
\begin{equation}\label{a16}
E(\xi,\sigma, n,t)=\begin{cases}
   t\sqrt{2 v_{F}^{2} \hbar e B |n|+\Delta_{\xi\sigma}^{2}}, & \text{if $n\neq 0$}.\\
    -\xi \Delta_{\xi\sigma}, & \text{if $n=0$}.
  \end{cases}
\end{equation}
Here, $t=\textrm{sgn(n)}$ denotes the conduction/valence band, $\Delta_{\xi\sigma}=-\frac{1}{2}\xi \sigma \Delta_{so}+\frac{1}{2}\Delta_{z}$ and $n$ is an integer, the quantum number denoting Landau quantization and $\sigma=\pm 1$ for spin up ($\uparrow$) and down  ($\downarrow$) respectively. Note that the $n$=0 manifold is independent of the magnetic field and these levels can be linearly manipulated by the electric field only, whereas both the electric and magnetic fields play a role in setting the position of the $n$$\neq$0 levels. The corresponding eigenfunctions at the $K$ and $K'$ points are
\begin{align}\label{a17}
|\bar n\rangle \bigg\rvert_{\xi=1}=\begin{pmatrix} -iA_{n} |n-1\rangle
  \\  B_{n} |n\rangle \\ \end{pmatrix}
  \end{align}
and
\begin{align}\label{a17}
|\bar n\rangle \bigg\rvert_{\xi=-1}=\begin{pmatrix} -iA_{n} |n\rangle
  \\  B_{n} |n-1\rangle \\ \end{pmatrix}
\end{align}
where $|n\rangle$ is an orthonormal Fock state of the harmonic oscillator, and $A_{n}$ and $B_{n}$ are given by,
\begin{align}\label{a18}
  A_{n} =& \begin{cases}
   \frac{\sqrt{|E(\xi,\sigma, n,t)|+t\Delta_{\xi\sigma}}}{\sqrt{2|E(\xi,\sigma, n,t)|}}, & \text{if $n\neq 0$}.\\
    \frac{1-\xi}{2}, & \text{if $n=0$}.
  \end{cases}
    \end{align}
and

  \begin{align}
  B_{n}= & \begin{cases}
   \frac{\sqrt{|E(\xi,\sigma, n,t)|-t\Delta_{\xi\sigma}}}{\sqrt{2|E(\xi,\sigma, n,t)|}}, & \text{if $n\neq 0$}.\\
    \frac{1+\xi}{2}, & \text{if $n=0$}.
  \end{cases}
\end{align}

The next task is to calculate the magneto-optical conductivity for the 2D systems admitting the Hamiltonian in Eq.~(\ref{a1}). As the authors in \cite{Tabert12013} describe, Kubo formula is used to derive the following general expressions for the conductivity~\cite{Tabert2013,Lasia2014},
\begin{equation}\label{c13}
\sigma_{\mu\nu}(\Omega)
=\frac{i\hbar}{2\pi l_{B}^{2}}\sum_{\sigma, \xi=\pm 1}\sum_{mn}\frac{f_{n}-f_{m}}{E_{n}-E_{m}}\thickspace
\frac{\langle \bar{n}|\hat{j}_{\mu}|\bar{m}\rangle\langle \bar{m}|\hat{j}_{\nu}|\bar{n}\rangle}{\hbar \Omega-(E_{n}-E_{m})+i\Gamma},
\end{equation}
where $f_{n}=1/(1+e^{{(E_{n}-\mu_{F})}/{k_{B}T}})$ is the Fermi Dirac distribution function at temperature $T$ and chemical potential $\mu_{F}$, $\hat{j}_{\mu}=ev_{F}\hat{s}_{\mu}$ is the current  operator, $\hat{s}_{\mu}$ are Pauli matrices, $E_{n}$ is the energy of the $n$'th landau
level, $\Gamma$ is the transport scattering rate responsible for the broadening of the energy levels and $l_{B}=\sqrt{\hbar/eB}$ is the magnetic length. The spatial index $\mu$ can be $x$, $y$ or $z$. The conductivity is complex whose real and imaginary parts can also be separately computed. For example, at $T=0$ K we have the longitudinal conductivity
\begin{widetext}
\begin{multline}\label{c19}
\frac{\textrm{${\textrm{Re} \above 0pt \textrm{Im}}$}\bigg\} \big(\sigma_{xx}(\Omega)\big)}{\sigma_{0}}=\frac{2 v^{2}\hbar e B}{\pi}\sum_{\xi, \sigma}\sum_{m,n}\frac{\Theta(E_{n}-\mu_{F})-\Theta(E_{m}-\mu_{F})}{E_{n}-E_{m}}\\\times\bigg[(A_{m}B_{n})^{2}
\delta_{|m|-\xi,|n|}+(B_{m}A_{n})^{2}\delta_{|m|+\xi,|n|}\bigg]\bigg\{\textrm{${F \above 0pt G}$},
\end{multline}
where, $\sigma_{0}=e^2/4\hbar$, $F=\Gamma/\bigg((\hbar\Omega-(E_n-E_m))^{2}+\Gamma^{2}\bigg)$ and $G=\big(\hbar\Omega-(E_n-E_m)\big)/\bigg((\hbar\Omega-(E_n-E_m))^{2}+\Gamma^{2}\bigg)$. In these expressions, the Kronecker deltas ensure the rules for electric dipole transitions between the LL's are satisfied.
The Heaviside functions $\Theta(E_{n}-\mu_{F})$ ensure that transitions across the Fermi level are possible, hence they effectively account for the so called Pauli blocking \cite{Grigorenko2012}. Similarly, the real and imaginary parts of the transverse conductivity are
\begin{multline}\label{c21}
\frac{\textrm{${\textrm{Re} \above 0pt \textrm{Im}}$}\bigg\} \big(\sigma_{xy}(\Omega)\big)}{\sigma_{0}}=\frac{2 v^{2}\hbar e B}{\pi}\sum_{\xi, \sigma}\sum_{m,n}\xi\frac{\Theta(E_{n}-\mu_{F})-\Theta(E_{m}-\mu_{F})}{E_{n}-E_{m}}\\\times\bigg[(A_{m}B_{n})^{2}
\delta_{|m|-\xi,|n|}-(B_{m}A_{n})^{2}\delta_{|m|+\xi,|n|}\bigg]\bigg\{\textrm{${-G \above 0pt F}$}\cdot
\end{multline}
\end{widetext}
In the limit $\Delta_{so}=\Delta_{z}=0$, we recover graphene's Hall conductivity \cite{Gusynin2005}. For these expressions, the real (imaginary) part of $\sigma_{xx} (\sigma_{xy})$ is a sum of absorptive Lorentzians, each of whose FWHM depends on the scattering rate $\Gamma$, higher $\Gamma$ resulting in broader and shorter peaks. Plots of these conductivities can be seen in previous works \cite{Tabert2013,Tabert12013}, which set the stage and provide the formalism for computing the magneto-optic rotations discussed in the present work. Likewise, the real (imaginary) part of $\sigma_{xy} (\sigma_{xx})$ is a sum of dispersive Lorentzians. These peaks are positioned at $\hbar\Omega=(E_{n}-E_{m})$, which we call the magneto-excitation energies. The transitions obey the appropriate selection rules namely $|n|-|m|=\pm1$ and the conservation of real spin implying that transitions between $\sigma=+1$ and $-1$ levels are spin forbidden.

\subsection{Magneto-Optical Rotations and Ellipticities}
We now present a general method to calculate the Faraday and Kerr rotation angles and the resulting ellipticities. Due to the rich LL structure, these MO effects are modulated by myriad stimuli such as electric and magnetic fields \cite{Crassee2011,Dolatabady2019,Poumirol2017}, chemical potential gating \cite{Dolatabady2019}, modification through doping, optical pumping \cite{Grebenchukov2019} as well as temperature \cite{Okada2016} and the substrate effect \cite{Crassee2011}.

Throughout this article, we consider a well-collimated, monochromatic, Gaussian beam of light with nontotal reflection impinging from one medium to the planar interface of the silicene-substrate system at an incidence angle $\theta_{1}$. The beam of light of frequency $\Omega$ has polarization in an arbitrary direction, and is propagating through the incident and transmitted materials with \emph{relative} permittivity and permeability $\varepsilon_n$ and $\mu_n$ respectively, where $n=\textrm{(1,2)}$. The beam make an angle $\theta_{2}$ in the substrate which is assumed to be semi-infinite, obviating the need to consider finite substrate size effects and thin-film interference \cite{Yang2013}. The wave vectors are $k_1$ and $k_2$, $k_{n}=\Omega\sqrt{\mu_{n}\varepsilon_{n}}$, $Z_n=Z_0\sqrt{\mu_{n}/\varepsilon_{n}}$ and $Z_0=\sqrt{\mu_{0}/\varepsilon_{0}}$, where $\mu_{0}$ and $\varepsilon_{0}$ are the vacuum permeability and permittivity respectively.
The Fresnel coefficients have been derived in previous work \cite{Kort-Kamp2015,Wu12017}:
\begin{eqnarray}\label{da5}
  r_{pp} &=& \frac{\alpha_{+}^{T}\alpha_{-}^{L}+\beta}{\alpha_{+}^{T}\alpha_{+}^{L}+\beta}, \\
  \label{da6}
  r_{ss} &=& -\bigg(\frac{\alpha_{-}^{T}\alpha_{+}^{L}+\beta}{\alpha_{+}^{T}\alpha_{+}^{L}+\beta}\bigg), \\
  \label{da7}
   t_{pp} &=& 2\frac{Z_{2}\varepsilon_{2}}{Z_{1}}\frac{k_{1z}\alpha_{+}^{T}}{\alpha_{+}^{T}\alpha_{+}^{L}+\beta}, \\
   \label{da8}
t_{ss} &=& 2\mu_{2}\frac{k_{1z}\alpha_{+}^{L}}{\alpha_{+}^{T}\alpha_{+}^{L}+\beta},
\end{eqnarray}
\begin{equation}
\label{da9}
 r_{sp}=t_{sp}=\frac{-2Z_{0}^{2}\mu_{0}\mu_{1}\mu_{2}k_{1z}k_{2z}(\sigma_{H}+\sigma_{xy}^{sym})}{Z_{1}(\alpha_{+}^{T}\alpha_{+}^{L}+\beta)},
 \end{equation}
 \begin{equation}\label{da10}
 r_{ps}=-\frac{k_{1}k_{2z}}{k_{2}k_{1z}}t_{ps}= 2\frac{Z_{0}^{2}\mu_{1}\mu_{2}}{Z_{1}}\frac{k_{1z}k_{2z}(\sigma_{xy}^{sym}-\sigma_{H})}{\alpha_{+}^{T}\alpha_{+}^{L}+\beta},
\end{equation}
where,
\begin{eqnarray}\label{da25}
  \alpha_{\pm}^{L} &=& (k_{1z}\varepsilon_{2}\pm k_{2z}\varepsilon_{1}+k_{1z}k_{2z}\sigma_{L}/(\varepsilon_{0}\Omega)), \\
\label{da26}
  \alpha_{\pm}^{T} &=& (k_{2z}\mu_{1}\pm k_{1z}\mu_{2}+\mu_{0}\mu_{1}\mu_{2}\sigma_{T}\Omega), \\
\label{da27}
   \beta &=& Z_{0}^{2}\mu_{1}\mu_{2}k_{1z}k_{2z}[\sigma_{H}^{2}-(\sigma_{xy}^{sym})^{2}] \cdot
\end{eqnarray}
Here, $k_{1z}=k_{1}\cos(\theta_{1})$ and $k_{2z}=k_{2}\cos(\theta_{2})$. The conductivities $\sigma_{L}(\sigma_{T})$ are the longitudinal (transverse) components. For homogeneous, isotropic media, $\sigma_{L}=\sigma_{T}=\sigma_{xx}=\sigma_{yy}$. The cross conductivity of a 2D system in the presence of magnetic field is antisymmetric \cite{Gusynin2006,Oliva-Leyva2017} $\sigma_{xy}=-\sigma_{yx}$. In fact, the cross conductivity $\sigma_{xy}$ has symmetric $\sigma_{xy}^{sym}$ and asymmetric $\sigma_{xy}^{antisym}$ parts. For anisotropic materials, such as phosphorene \cite{Low2014}, $\sigma_{xy}^{sym}$ is non-zero because the band structure of phosphorene is Dirac like (linear in $k$) in one direction and Schrodinger like (parabolic in $k$) in the other direction \cite{Low12014}. However, for isotropic materials such as graphene and other staggered materials (silicene, germanene, stanene, and plumbene etc.) $\sigma_{xy}^{sym}$=0. Therefore in Eqs.~(\ref{da9}) and (\ref{da10}), we use $\sigma_{H}=\sigma_{xy}$ which comprises wholly of the anti-symmetric part. 

In our case, medium 1 is vacuum ($\varepsilon_{1}$= 1, $\mu_{1}$=1) and medium 2 is nonmagnetic $\mu_{2}$=1. The Fresnel coefficients which are derived from the magneto-optical conductivities, subsequently determine the magneto-optic rotations and ellipticity.
For incident $s$ and $p$ polarization, the Faraday rotation and ellipticity are computed using the expressions
\begin{eqnarray}\label{g1}
 \Theta^{\textrm{F},\textrm{s}(\textrm{p})} &=& \frac{1}{2}\tan^{-1}\bigg(2\frac{\textrm{Re} \big(\chi^{\textrm{F},\textrm{s}(\textrm{p})}\big)}{1-|\chi^{\textrm{F},\textrm{s}(\textrm{p})}|^{2}}\bigg), \\
 \text{and}\thickspace\thickspace
 \label{g2}
\eta^{\textrm{F},\textrm{s}(\textrm{p})}&=& \frac{1}{2}\sin^{-1}\bigg(2\frac{\textrm{Im} \big(\chi^{\textrm{F},\textrm{s}(\textrm{p})}\big)}{1-|\chi^{\textrm{F},\textrm{s}(\textrm{p})}|^{2}}\bigg),
\end{eqnarray}
where,
\begin{eqnarray}\label{g12}
 \chi^{\textrm{F},\textrm{s}}=\frac{t_{ps}}{t_{ss}} &=& Z_{0}\sqrt{\frac{\varepsilon_{1}}{\mu_{1}}}\frac{k_{1}cos(\theta_{1})\sigma_{H}}{\alpha_{+}^{L}}, \\
\text{and}\thickspace\thickspace
 \label{g22}
\chi^{\textrm{F},\textrm{p}}=\frac{t_{sp}}{t_{pp}}&=& -Z_{0}\sqrt{\frac{\mu_{2}}{\varepsilon_{2}}}\mu_{0}\mu_{1}\frac{k_{2}cos(\theta_{2})\sigma_{H}}{\alpha_{+}^{T}}\cdot
\end{eqnarray}
Similarly, for MOKE, the rotations and ellipticities are
\begin{eqnarray}\label{g3}
 \Theta^{\textrm{K},\textrm{s}(\textrm{p})} &=& \frac{1}{2}\tan^{-1}\bigg(2\frac{\textrm{Re} \big(\chi^{\textrm{K},\textrm{s}(\textrm{p})}\big)}{1-|\chi^{\textrm{K},\textrm{s}(\textrm{p})}|^{2}}\bigg), \\
\text{and}\thickspace\thickspace\eta^{\textrm{K},\textrm{s}(\textrm{p})}&=& \frac{1}{2}\sin^{-1}\bigg(2\frac{\textrm{Im} \big(\chi^{\textrm{K},\textrm{s}(\textrm{p})}\big)}{1-|\chi^{\textrm{K},\textrm{s}(\textrm{p})}|^{2}}\bigg),
\end{eqnarray}
where,
\begin{eqnarray}\label{g32}
\chi^{\textrm{K},\textrm{s}}=\frac{r_{ps}}{r_{ss}} &=& \frac{2Z_{0}\sqrt{\mu_{1}\varepsilon_{1}}\mu_{2}k_{1z}k_{2z}\sigma_{H}}{\alpha_{-}^{T}\alpha_{+}^{L}+\beta}, \\
\text{and}\thickspace\thickspace
 \label{g33}
\chi^{\textrm{K},\textrm{p}}=\frac{r_{sp}}{r_{pp}}&=& \frac{-2Z_{0}\sqrt{\mu_{1}\varepsilon_{1}}\mu_{0}\mu_{2}k_{1z}k_{2z}\sigma_{H}}{\alpha_{+}^{L}\alpha_{-}^{L}+\beta}\cdot
\end{eqnarray}
A note about the notation is in place here. The spin ($\uparrow$ or $\downarrow$) or  valley ($K$ or $K'$) will be specified in the subscripts while the superscripts identify the Faraday (F) or Kerr rotation (K) as well as the polarization state ($s$) or ($p$). If the $\chi$'s are small, $\chi\ll1$, Eqs.~(\ref{g1}) and (\ref{g2}) reduce to $\Theta^{\textrm{F},\textrm{s}(\textrm{p})}\approx\textrm{Re} (\chi^{\textrm{F},\textrm{s}(\textrm{p})})$ and $\eta^{\textrm{F},\textrm{s}(\textrm{p})}\approx\textrm{Im}(\chi^{\textrm{F},\textrm{s}(\textrm{p})})$ and vis-a-vis for the Kerr effect. However, for Landau quantized systems, there is no reason to believe, at the ontset, that the MO effects are small.

\section{Results and Discussion}

\begin{figure*}[!t]
   \includegraphics[scale=1.0]{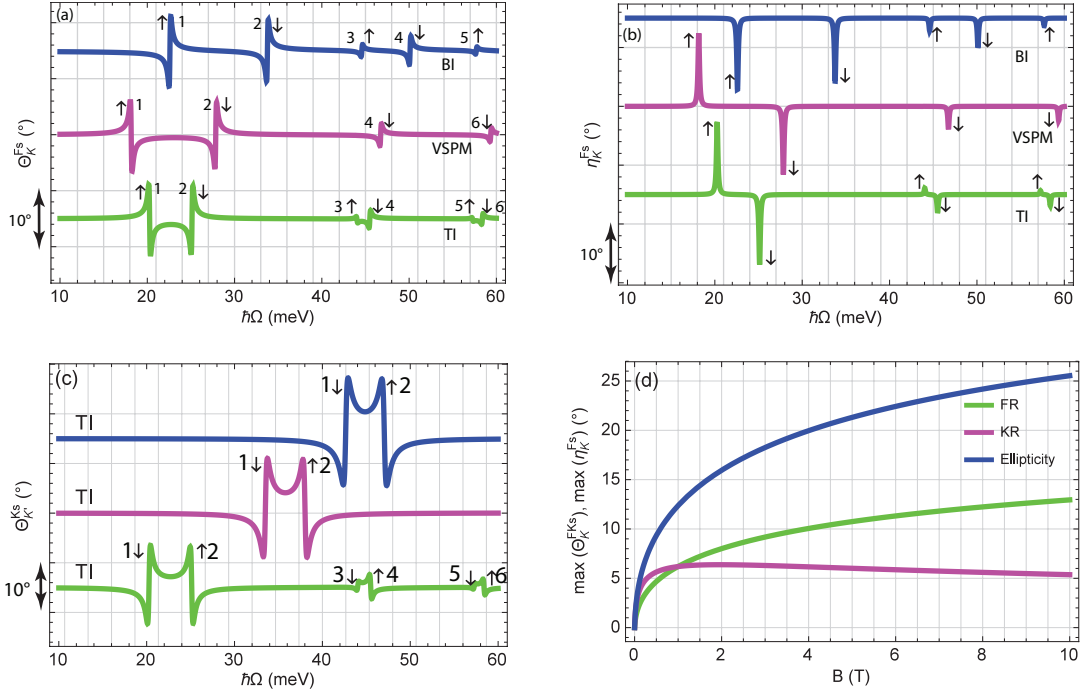}
 \caption{(Color online) Faraday, Kerr rotation and ellipticity of silicene-substrate system as function of photon energy electric and magnetic fields. (a) The $s$ polarized Faraday rotation and (b) ellipticity as function of incident photon energy in the $K$ valley with modulation of the external electric field for the three distinct topological regimes, TI, VSPM and BI for a magnetic field of $1$ T. The spectral peaks are labelled 1 through 6 and their origin is identified in the main text. The spectrum are vertically shifted by $15^{\circ}$ among themselves for clearer viewing. Furthermore, in this figure we use $\Delta_{z}=\Delta_{so}/2$ (TI) and $\Delta_{z}=2\Delta_{so}$ (BI). (c) The $s$ polarized Kerr rotation as function of incident photon energy in the $K'$ valley with modulation of the external magnetic field for the TI regime for three different values of $B=$1, 3 and 5 T. (d) The maximum Faraday, Kerr rotation and ellipticity as function of magnetic field  in $K$ valley for the single transition $\Delta_{-10,K, \uparrow}$. The parameters used are $\theta_{1}=30^{\circ}$, $\Gamma=0.01\Delta_{so}$, refractive index $n_{2}=3.4$ and chemical potential $\mu_{F}=0$.}
\label{fig1}
\end{figure*}

First we discuss Faraday rotation (FR) for charge neutral 2D silicene, where the inter-band transitions bridge across the valance and conduction bands. Hence $\mu_{F}=0$. Fig~\ref{fig1}(a) shows the FR spectra as a function of incident photon frequency with modulation of the external electric field, landing the band structure into three distinct topological regimes \cite{Tabert2013}. The signal originating from a single spin orientation in only one of the valleys is dispersive Lorentzian, with a positive followed by a negative (or vice versa) signature. Let's call this an anti-phase peak. This terminology is borrowed from NMR literature \cite{Levitt2008}. The anti-phase peak is centred at the magneto-optic excitation frequency $E_{n}-E_{m}$ with positive and negative maxima at $E_{n}-E_{m}\pm\Gamma$. For the opposite spin in the same valley and an identical LL transition, we still see an absorptive anti-phase peak whose sign may be reversed, the possibility of reversal depending on the exact topological regime. The peaks corresponding to the different transitions, $E_{m,K (K'),\uparrow(\downarrow)}\rightarrow E_{n,K (K'),\uparrow(\downarrow)}$ are labelled as $\Delta_{mn,K (K'),\uparrow(\downarrow)}$. For higher frequencies, the magnitude of the rotation is reduced in accordance with the factor of $1/(E_{n}-E_{m})$ appearing in the denominator of Eqs.~(\ref{c19}) and (\ref{c21}). We now explore the three distinct topological regimes.

In the topological insulator (TI) regime $(\Delta_{z} < \Delta_{so})$, the first and second anti-phase peaks correspond to the $\Delta_{-10,K, \uparrow}$ and $\Delta_{01,K, \downarrow}$ transitions for spin up and spin down respectively. In each of these transitions, one of the participating levels is an $n$=0 level.  In a magnetic field of $1$ T and $\Delta_{z}=\Delta_{so}/2$, these magneto-excitation energies are calculated as 20.3 meV (4.9 THz) and 25.1 meV (6.1 THz) respectively and are shown as 1 and 2 in the bottom spectrum of Fig.~\ref{fig1}(a). The anti-phase peaks switch sign with spin within the same valley. The $s$ polarized FR angles for the first two anti-phase peaks are $\sim \pm 6.5^{\circ}$. The subsequent anti-phase peaks appearing at different resonant frequencies differ in magnitude for spin up and spin down cases due to spin dependent energies. The anti-phase peaks labelled 3 through 6 can also be assigned to the various transitions. For example the multiplet structure 3 originates from $\Delta_{-12,K, \uparrow}$, 4 is due to $\Delta_{-21,K, \downarrow}$, 5 is due to $\Delta_{-23,K, \uparrow}$ and 6 comes from $\Delta_{-32,K, \downarrow}$.

In the valley-spin polarized metal (VSPM) instance $(\Delta_{z}=\Delta_{so})$, the gap of one of the spin-split bands closes \cite{Tabert2013} giving rise to a Dirac point. As we increase the applied electric field and begin to approach the VSPM point, the lowest frequency peaks, labeled 1 and 2 in the middle spectrum of Fig.~\ref{fig1}(a) move apart: the $\Delta_{-10,K, \uparrow}$ peak is red shifted and $\Delta_{01,K, \downarrow}$ peak is blue shifted. The excitation energies corresponding to the first two anti-phase peaks at the VSPM point are now 18.2 meV (4.4 THz) and 27.8 meV (6.7 THz). However, it is observed that at this precise electric field, the spectrum is cleaner and exhibiting fewer peaks. The peaks labeled 4 and 6 originate from the $\Delta_{-21,K, \downarrow}$  and $\Delta_{-32,K, \downarrow}$ transitions. The peaks that were labelled 3 and 5 in the TI regime and came from the $\Delta_{-12,K, \uparrow}$ and $\Delta_{-23,K,\uparrow}$ transitions are now annihilated. Eq.~(\ref{a16}) shows that at the Dirac point in the $K$ valley ($\xi=1$), the spin up ($\sigma=1$) transitions leads to $\Delta_{\xi\sigma}=0$ which results in $A_{n}=B_{n}=1/\sqrt{2}$ irrespective of the Landau quantum number $n$. For these spin up levels, therefore $(A_{m}B_{n})^{2}=(B_{m}A_{n})^{2}$ and from Eq.~(\ref{c21}), the minus sign between the terms in the square brackets results in annihilation of the spectral response at the $\Delta_{-12}$  and $\Delta_{-23}$ frequencies. So even though, these transitions are allowed by selection rules, destructive interference between their quantum amplitudes extinguishes the response. Conversely, in the $K'$ valley (data not shown), the spin down peaks will be annihilated at the Dirac points.

For an even higher electric field $(\Delta_{z} > \Delta_{so})$, the system transitions from the VSPM to the band insulator (BI) state and the lowest band gap is opened again, resulting in sign change of some of the anti-phase peaks with respect to the TI phase. Compare the peaks 1 through 5 between the TI and BI shown in Fig.~\ref{fig1}(a). The full range of the allowed peaks also resurfaces once the VSPM point is crossed. The separation between the anti-phase pair keeps on growing in the BI state. Consequently, all the peaks gradually shift towards higher frequencies. The magnitude of the maximum spin polarized FR angles for the first two peaks inside the anti-phase pair is $\sim\pm 8^{\circ}$.  If we change the polarization of the incident light, the sign of anti-phase peaks inverts with respect to the baseline.  Alternatively the same effect is achieved by switching from one valley to another. If we change the valley, the spin identity of the anti-phase also changes. The juxtaposition of identities between spin up and down polarized peaks after band inversion is also observed in the $K'$ valley. Consequently the $s$ polarized FR in the $K$ valley will have the same form as the $p$ polarized FR response in the $K'$ valley. The MOKE rotation spectra (data not shown) follow a similar trend. The MOKE response is also spin and valley polarized and the magnitudes of the rotation angles range between $5\textendash15^{\circ}$ for both valleys and all three topological regimes, which are in general larger than the FR angle.
\begin{figure*}[!t]
   \includegraphics[scale=1.0]{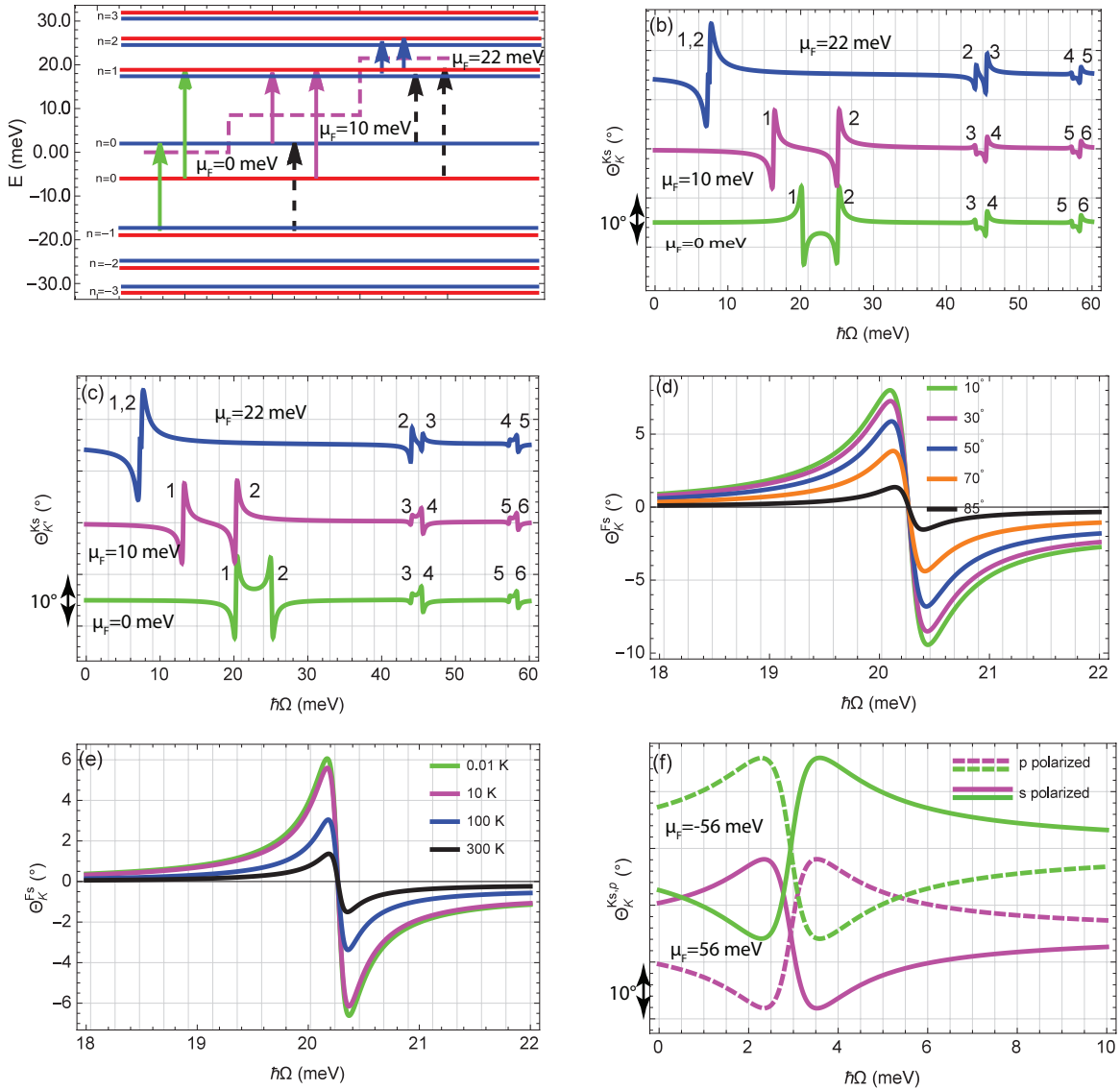}
 \caption{(Color online) (a) Schematic representation of the allowed transitions
between LL's for three different values of chemical potential $\mu_{F}=$0, 10 and 22 meV; (b) and (c) the $s$ polarized Kerr
rotation as function of incident photon frequency in $K$ and $K'$ valleys with modulation of the chemical potential in the TI regime for a magnetic field of 1 T, respectively. (d) The $s$ polarized Faraday rotation as function of incident photon energy in $K$ valley for different incident angles for a single transition in the TI regime. (e) The $s$ polarized Faraday rotation as function of incident photon energy in $K$ for different temperatures for a single transition in the TI regime. (f) The $s$ and $p$ polarized Kerr rotation as function of incident photon frequency in the semiclassical limit for n-type and p-type silicene ($\mu_{F}=$ 56 and -56 meV), respectively. The solid line represents the $s$ polarized and the dashed line $p$ polarized. The parameters used are $\theta=30^{\circ}$, $\Gamma=0.01\Delta_{so}$ and refractive index $n_{2}=3.4$.}
\label{fig2}
\end{figure*}

Fig.~\ref{fig1}(b) shows the series of peaks in the ellipticity acquired by transmitted light from $s$ polarized incident radiation originating from the $K$ valley manifold. Faraday geometry is considered though analogous results are obtained for reflection as well. It is evident that extermely large ellipticities, of the order of $8$\textendash$15^{\circ}$, appear for the lowest excitations. The spectrum for ellipticity comprises absorptive Lorentzians, which are spin and valley polarized. These maxima are at the excitation energies $E_{n}-E_{m}$. The rotation and ellipticity data when considered together, indicate that at the exact excitation energy $E_{n}-E_{m}$, the rotation is zero while the ellipticity is maximum. Furthermore, when the rotation is maximum $(\hbar\Omega=E_{n}-E_{m}\pm\Gamma)$, the ellipticity drops to $50\%$ of its maximum value. The intertwined effects, although both being ultra-large, limit the use of silicene-substrate system for a \emph{pure} MO rotator since significant ellipticity is also introduced.

We now demonstrate the effect of how the magnetic field modifies the magneto-optic response. The $s$ polarized MOKE in the $K'$ valley is only one possible illustration and shown in Fig.~\ref{fig1}(c). Here we plot the MO spectrum in the TI regime for three different values of $B=$1, 3 and 5 T, while keeping $\mu_{F}=0$ and $\theta=30^{\circ}$. The impact on the MOKE signal in terms of shifting magneto-excitation frequency and the amount of Kerr rotation is clear. The silicene energy levels are strongly dependent on the magnetic field $B$, as given by Eq.~(\ref{a16}), and this is also true for other 2D materials including graphene \cite{Koppens2011,Tabert2013}. As we increase the strength of the applied magnetic field, the MO excitations shift towards higher frequencies with a concomitant increase in magnitude of the MOKE rotation angle. For example, the peaks labelled 1 and 2 have excitation energies 20.3 meV (4.9 THz) and 25.1 meV (6.1 THz) for $B=1$ T, 33.5 meV (8.1 THz) and 38 meV (9.2 THz) for $B=3$ T and finally, 42.7 meV (10.3 THz) and 47 meV (11.32 THz) for $B=5$ T. The maximum value of the rotation $\Theta_{K'}^{Ks}$ exceeds $\sim \pm 13^{\circ}$ at a magnetic field of 5 T, which is an exceptionally large rotation for a monolayer silicene-substrate system. Similarly the FR is also strongly field-dependent (data not shown). The primary role of the magnetic field tuning, therefore, is to shift the position of the magneto-optic excitation energies and also to modify the amount of rotation. However, unlike the electric field the magnetic field does not switch the sign of the anti-phase doublets.

Fig.~\ref{fig1}(d) show the field dependence of $\Theta_{K}^{Fs}$ and $\Theta_{K}^{Ks}$ in the TI regime. Due to the dispersive MO spectrum, we chose to plot the maximum rotation. By increasing the magnetic field strength the amount of FR and Kerr rotations grows.  However for stronger fields, the Kerr signal slowly decreases. At a field of $10$ T, we report $\Theta_{K}^{Fs}=13^{\circ}$ and $\Theta_{K}^{Ks}=5.5^{\circ}$. The ellipticity is also strongly field dependent.

It is also instructive to discuss the effect of controlling the FR and MOKE spectra by varying the chemical potential of the silicene surface, e.g, by applying a bias voltage \cite{Dolatabady2019} or optical pumping \cite{Grebenchukov2019}. For illustration purposes, we consider three different values of chemical potentials $\mu_{F}$=0, 10 and 22 meV,  while keeping the magnetic field 1 T, in the TI regime ($\Delta_{z}=0.5\Delta_{so}$) and an angle of incidence of $30^{\circ}$. In the first case the chemical potential is at zero and lies within the $n$=0 manifold, for $\mu_{F}$=10 meV the chemical potential is in between the $n$=0 and $n$=1 LL's and for $\mu_{F}$=22 meV the chemical potential is in between the $n$=1 and $n$=2 LL's. These LLs are shown in Fig.~\ref{fig2}(a). Only the $K$ valley is depicted. For $\mu_{F}=0$, the lowest energy excitations are also indicated on the same subfigure. They identify as $\Delta_{-10,K,\uparrow}=20.3$ meV and $\Delta_{01,K,\downarrow}=25.1$ meV. These result in the Kerr rotations shown by anti-phase peaks 1 and 2 in the bottom spectrum of Fig.~\ref{fig2}(b). They are interband transitions since they occur across the zero energy datum. The combination of anti-phase peaks $\Delta_{-12}$ and $\Delta_{-21}$ in the $K$ valley, for both spins yield the multiplet structure 3 and the transitions $\Delta_{-23}$, $\Delta_{-32}$ yield the structure 4. These anti-phase are rather close in their excitation energies (e.g. $\Delta_{-12,K,\uparrow}=44.0$ meV, $\Delta_{-21,K,\uparrow}=44.0$ meV, $\Delta_{-12,K,\downarrow}=45.5$ meV and $\Delta_{-21,K,\downarrow}=45.5$ meV) and the ability to resolve this finer structure depends on the experimental capability.

As $\mu_{F}$ increases to $10$ meV, certain transitions become Pauli blocked. For example the transition $\Delta_{-10,K,\uparrow}$ becomes forbidden and in its stead, the intra-band transition $\Delta_{01,K,\uparrow}=16.0$ meV emerges. The Pauli blocked transition is shown by a dashed upward pointing arrow in the middle part of Fig.~\ref{fig2}(a) and the two lowest transitions, $\Delta_{-10,K,\uparrow}=16.3$ meV and $\Delta_{01,K,\downarrow}=25.1$ meV are shown by solid arrows. Once again, these yield the Kerr signatures 1 and 2 shown in Fig.~\ref{fig2}(b). The higher frequency agglomerated multiplets 3 and 4 remain unchanged. If $\mu_{F}$ is further increased to $22$ meV, so that it lies between the $n$=1 and $n$=2 manifolds, both transitions starting from $n$=0, i.e,  $\Delta_{-10,K,\uparrow}$ and $\Delta_{01,K,\downarrow}$ now become Pauli blocked. These are again indicated by the dashed arrows in the rightmost part of Fig.~\ref{fig2}(a). In their place, however, the intra-band transitions $\Delta_{12,K,\uparrow}=7.0$ meV and $\Delta_{12,K,\downarrow}=7.2$ meV pop up. For higher $n$, the LL's are closely spaced. Hence the excitation energies also converge. These closely spaced transitions are separated by $ \approx 200~\mu$eV and are shown by the structure 1, 2 in top part of Fig.~\ref{fig2}(b). The transitions involving the $n$=0 levels are completely missing from this magneto-optic spectrum. Furthermore, the transitions $\Delta_{-21,K,\uparrow}$ and $\Delta_{-21,K,\downarrow}$ become Pauli forbidden and hence are absent from the excitation structure labeled 3 which now comprises only $\Delta_{-12,K,\uparrow}=44.0$ meV and $\Delta_{-12,k,\downarrow}=45.5$ meV. Therefore peak 3 is a cleaner doublet of anti-phase structure when compared with the $\mu_{F}$=0 and $\mu_{F}$=10 meV cases. The structure 4 originates, as earlier, from rather  closely spaced $\Delta_{-23}$ and $\Delta_{-32}$.

The magneto-optic spectrum originating from the $K'$ valley for the same values of $\mu_{F}$ is depicted in Fig.~\ref{fig2}(c). For $\mu_{F}=0$, the rotational peaks are coincident with the $K$ valley as $\Delta_{mn,K,\uparrow}=\Delta_{nm,K',\downarrow}$ (when $m=0$ and $n\neq0$), $\Delta_{-10,K,\uparrow}=\Delta_{01,K',\downarrow}$ and $\Delta_{01,K,\downarrow}=\Delta_{-10,K',\uparrow}$. However, these valley-specific spectrums are sign inverted with respect to each other. For $\mu_{F}=10$ meV, the lowest energy transitions 1 and 2 occur at different positions for the two valleys. For the $K$ valley, 1 and 2 are $\Delta_{01,K,\uparrow}=16$ meV and $\Delta_{01,K',\downarrow}=25$ meV respectively whereas for the $K'$ valley, the peaks 1 and 2 are $\Delta_{01,K',\uparrow}=13.1$ meV and $\Delta_{01,K',\downarrow}=20.3$ meV respectively.

The FR and MOKE signatures are clearly sensitive to the incident angle $\theta_{1}$. This is because the Fresnel coefficients are strongly dependent on the incidence angle. This dependence is shown in Fig.~\ref{fig2}(d) for a single transition in the TI regime. An increasing incidence angle diminishes the amount of rotation until it disappears at complete grazing, $\theta_{1}=\pi/2$. A similar trend can also be seen in the MO response of graphene \cite{Xu2013,Yoshino2013}.

All of the results presented so far are at $0$ K but as the temperature goes up, the Fermi Dirac distribution function in Eq.~(\ref{c13}) starts becoming significant. We can explore the temperature dependence of the FR by introducing these distributions in place of the Heaviside functions. Nevertheless, at $100$ K, the FR angle is $3^{o}$, while at $300$ K, the FR angle is $1^{\circ}$. These results are shown in Fig.~\ref{fig2}(e). The experimental value \cite{Okada2016} of the FR angle for graphene at $1.5$ K is $4$ mrad which translates to $0.23^{\circ}$. This shows that silicene has a bigger Faraday rotation than graphene in the THz range. In silicene the electrons frequently interact with scatterers. There are many scattering mechanisms including Coulomb interaction, impurities, optical phonons, acoustic phonons, and radiative decay \cite{Oliva-Leyva2017}.  Due to these scattering channels the peaks are additionally broadened \cite{Funk 2015}. However, in actuality, the temperature dependence of the scattering rate $\Gamma$ must also be taken into account. This is ignored in the present work. 

\begin{table*}[]
\caption{Table of allowed transitions in $K$ valley in the $n=-1,0,1$ subspace, at a fixed magnetic field and chemical potential $\mu_{F}$. Furthermore $x=\Delta_{z}/\Delta_{so}$, $y=\sqrt{\hbar v^{2}eB/\Delta_{so}^{2}}.$ and $\mu=\mu_{F}/\Delta_{so}$.}
\centering
\begin{tabular}{lllll}
\hline \hline
$m$  & $n$  & spin ($\uparrow\downarrow$) & Range of $x$  & $\Delta_{mn, K(K'),\uparrow(\downarrow)}$ \\ \hline
0 & 1 & $\uparrow$ & $x\geq1-2\mu$ &~~~~~~$-\frac{1}{2}+\frac{x}{2}+\sqrt{(\frac{x-1}{2})^{2} +2y^{2}}$ \\
0 & 1 & $\downarrow$ &  all $x$ &~~~~~~$+\frac{1}{2}+\frac{x}{2}+\sqrt{(\frac{x+1}{2})^{2} +2y^{2}}$ \\
-1 & 0 & $\uparrow$ &  $x\leq1-2\mu$ &~~~~~~$+\frac{1}{2}-\frac{x}{2}+\sqrt{(\frac{x-1}{2})^{2} +2y^{2}}$ \\
-1 & 0 & $\downarrow$ & not allowed &~~~~~~ \\ \hline \hline
\end{tabular}
\end{table*}
 \begin{figure}[!h]
   \includegraphics[scale=1]{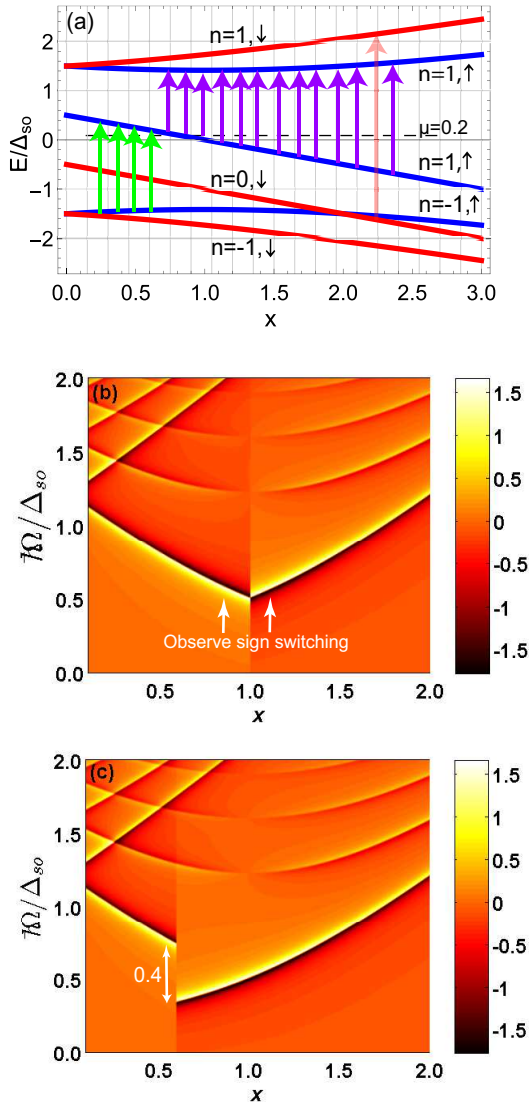}
 \caption{ (a) Schematic representation of the allowed transitions between LL's for chemical potential $\mu=0.2$. (b) and (c) the $s$ polarized Faraday rotation contour plots as function of $x$ in $K$ valley for $\mu=$ 0 and 1.25, repectively, where $x=\Delta_{z}/\Delta_{so}$ and $\mu=\mu_{F}/\Delta_{so}$. The parameters used are $\theta=30^{\circ}$, $\Gamma=0.01\Delta_{so}$ and refractive index $n_{2}=3.4$.}
  \label{fig3}
\end{figure}
In studies on 2D materials placed inside magnetic fields, the semiclassical limit is valid when the LL spacing becomes unimportant and inconsequential \cite{Tabert12013}. This happens as $|n|$ goes up and when the chemical potential is high up in the conduction band or deep down in the valance band, $|\mu_{F}|\gg|E_{0}|$. In this case the intra-band transitions between closely spaced levels are allowed. Suppose that $\mu_{F}$ lies between the $n-1$ and $n$ LL's. Since the gap is minuscule, $\mu_{F}\approx E_{n}$. In this limit we have
\begin{equation}\label{g5}
E_{n+1}-E_{n}\approx\frac{\hbar v^{2}e B}{\sqrt{\Delta_{\xi \sigma}^{2}+2n\hbar v^{2}e B}}=\hbar\Omega_{c},
\end{equation}
where, $\Omega_{c}=\hbar v^{2}e B/\mu_{F}$ is called the classical cyclotron frequency. In this regime the Faraday and Kerr rotations can also be derived from a purely classical point of view \cite{Tabert2013}. For finite $\mu_{F}$ and $n$, the allowed transitions are $\Delta_{(n-1)n}$, $\Delta_{-(n-1)n}$ and $\Delta_{-(n+1)n}$, however the latter two are large energies with diminished contributions to the magneto-optical conductivities, Eqs.~(\ref{c19}) and (\ref{c21}). Hence, the allowed transition is the one that immediately across the chemical potential and results in a \emph{single} large peak in all magneto-optic signatures. Furthermore in the semiclassical limit $A_{n}\approx A_{n-1}\approx B_{n}\approx B_{n-1}=1/\sqrt{2}$, and we can straightforwardly derive, using Eqs.~(\ref{c19}) and (\ref{c21}), the following conductivities summed over both valleys and both spins,
\begin{equation}\label{g8}
\frac{\textrm{Re}\big(\sigma_{xx}(\Omega)\big)}{\sigma_{0}}=-\frac{\textrm{Im}\big(\sigma_{xy}(\Omega)\big)}{\sigma_{0}}
=\frac{\hbar\mu_{F}}{\pi}\frac{\Gamma}{(\hbar(\Omega-\Omega_{c}))^{2}+\Gamma^{2}},
\end{equation}
\begin{equation}\label{g9}
\frac{\textrm{Im}\big(\sigma_{xx}(\Omega)\big)}{\sigma_{0}}=\frac{\textrm{Re}\big(\sigma_{xy}(\Omega)\big)}{\sigma_{0}}
=\frac{\hbar\mu_{F}}{\pi}\frac{\hbar(\Omega-\Omega_{c})}{(\hbar(\Omega-\Omega_{c}))^{2}+\Gamma^{2}}\cdot
\end{equation}

These conductivities are shaped as absorptive and dispersive Lorentzians and are directly used to compute the Fresnel coefficients Eqs.~(\ref{da5})\textendash(\ref{da10}) and subsequently the rotations. The conductivities, therefore, are modeled by classical Drude-like  behavior \cite{Crassee2011,Tymchenko2013}. For example, in Fig.~\ref{fig2}(f) we plot the $s$ and $p$ polarized Kerr rotation angles as a function of the incident photon frequency in the $K$ valley. For n-type doping, we set $\mu_{F}$=56 meV, which places chemical potential between the $n$=9 and $10$ LL's. The transitions from $n$=0 to higher LL's are Pauli blocked, the selection rules dictate that only three transitions $\Delta_{-9~10}$, $\Delta_{-11~10}$ and $\Delta_{9~10}$ are allowed. The former two have negligible contributions, whereas the last mentioned transition results in a strong Drude peak, at 2.94 meV (0.71 THz). Similarly we plotted the $s$ and $p$ polarized Kerr rotation angles as function of the incident photon frequency for p-type silicene, with $\mu_{F}=-56$ meV. The Kerr rotation angle switches between n-type and p-type silicene indicating modulation of the rotation angle by switching the chemical potential, e.g, by switching gate bias voltage. Also note that the spin and valley information is lost in the semiclassical limit.  The value of electric field $\Delta_{z}$ also becomes inconsequential at higher doping and the silicene behaves as graphene, because not only that the resonant frequency approaches that of graphene in this limit \cite{Tabert2013}, but also the role of SOI becomes inconsequential.

An alternative approach to understanding the magneto-optic response is by contour plotting the rotations as a function of two variables. This method also allows one to identify topologically distinct regions and topological phase transitions \cite{Tabert2013} and may  reveal discontinuations that may otherwise go unnoticed. For example, we consider transitions in the $K$ valley within the $n=-1,0,1$ subspace. We use dimensionless variables to simplify the analysis. We can define $x=\Delta_{z}/\Delta_{so}$, $y=\sqrt{\hbar v^{2}eB/\Delta_{so}^{2}}$ as measures of the electric and magnetic fields respectively, $\hbar\Omega/\Delta_{so}$ and $\mu=\mu_{F}/\Delta_{so}$ as variables for photon frequency and chemical potential. In Fig.~\ref{fig3}(a), we first plot the LL spectrum for the transitions under consideration. Table I summarizes the allowed transitions across the chemical potential. The excitation energies are also computed in the last column. It is evident that at the critical point $x=1-2\mu$, the $\Delta_{-10,K,\uparrow}$ transition gives way to the $\Delta_{01,K,\uparrow}$ transition, we say that the former becomes Pauli blocked. For example for a precise value of $\mu=0.2$, this is shown by the sequence of green colored arrows that are only drawn for $x\leq1-2\mu$ and the purple colored arrows drawn only for $x\geq1-2\mu$. For $\mu=0$, this transition point is $x=1.0$. The contour plot in Fig.~\ref{fig3}(b) aptly captures the scenario. For $x\leq1.0$, the $\Delta_{-01,K,\uparrow}$ transitions causes the Faraday rotation while for $x\geq1.0$, the $\Delta_{01,K,\uparrow}$ transition kicks in yielding the Faraday rotation.
Upon this transition point, the sign of the anti-phase peaks also switches. For $\mu=0.2$, this switching now occurs at a smaller value of $x_{c}=1-2\mu=0.6$ as depicted in Fig.~\ref{fig3}(c). Furthermore at this switching point, $x_c=1-2\mu$, one observes a discontinuous jump in the excitation energy. This can be computed by inserting $x_c$ into the energies $\Delta _{01,K,\uparrow}|_{x_c}=-\mu+\sqrt{\mu^2+2y^2}$ and $\Delta _{-10,K,\downarrow}|_{x_c}=\mu+\sqrt{\mu^2+2y^2}$ which yields a discontinuity of magnitude $2\mu$ in the contour plot. The uninterrupted rotation which continues upward to the top right is due to $\Delta_{01,K,\downarrow}$ transition which is always switched on and is shown by thick red colored arrows in Fig.~\ref{fig3}(a). The $x_c$ point also indicates a topological phase transition from the TI to the BI regime.

\section{CONCLUSIONS}
In conclusion, we have theoretically demonstrated the transitional MO effect due to the topological phase transition in silicene. We have studied the electric field modulated valley and spin polarized Faraday, Kerr rotations and ellipticities for three different topological regimes in silicene. We found that the magnitude of the maximum valley and spin polarized FR and MOKE angles for the first two anti-phase pair is $8^{\circ}$ and $13^{\circ}$, respectively. We also observe that if we change the polarization of the incident light or switched from one valley to another, the anti-phase peaks invert with respect to the baseline. We further investigated the magnetic field modulated MOKE for different magnetic fields and found that by increasing the magnetic field, the positions of the valley and spin polarized FR and MOKE anti-phase peaks move towards higher frequencies and the amount of FR and MOKE rotation is also enhanced. Moreover, we also note the effect of varying chemical potential on valley and spin polarized FR and MOKE.

\end{document}